\pdfoutput=1
\documentclass[final]{elsarticle}

\graphicspath{{.},{./imgs/}}
\DeclareGraphicsExtensions{.pdf,.jpeg,.png}
\usepackage{subfig}
\usepackage{xcolor}
\usepackage{algorithmic}
\usepackage[utf8]{inputenc}
\usepackage{url}
\usepackage{multirow}
\usepackage{verbatim}
\usepackage{amsmath}
\usepackage{dsfont}
\usepackage{tikz}

\journal{Future Generation Computer Systems (FGCS) - Special Issue on
Recent Developments in High Performance Computing and Security}
\begin{document}

\begin{frontmatter}
\title{Fast Computation of the Performance Evaluation of Biometric Systems: Application to Multibiometrics}

\author{Romain Giot\corref{corres}}
\ead{romain.giot@ensicaen.fr}
\author{Mohamad El-Abed}
\ead{mohamad.elabed@ensicaen.fr}
\author{Christophe Rosenberger}
\ead{christophe.rosenberger@ensicaen.fr}

\cortext[corres]{Corresponding author}
\address{GREYC Laboratory\\ ENSICAEN - University of Caen Basse Normandie -
CNRS\\ 6 Boulevard Mar\'echal Juin, 14000 Caen Cedex - France}

\begin{abstract}
{
The performance evaluation of biometric systems is a crucial step when designing and evaluating such systems.
The evaluation process uses the Equal Error Rate (EER) metric proposed by the
International Organization for Standardization (ISO/IEC).
The EER metric is a powerful metric which allows easily comparing and
evaluating biometric systems.
However, the computation time of the EER is, most of the time, very intensive.
%Biometric systems need to be evaluated during their design, or during their
%parameter configuration.
%One common functional point used for the evaluation is the Equal Error Rate.
%Depending on the quantity of involved scores, the EER computation may be very
%computational intensive.
In this paper, we propose a fast method which computes an approximated value of
the EER.
We illustrate the benefit of the proposed method on two applications:
the computing of non
parametric confidence intervals and the use of genetic algorithms to
compute the parameters of fusion functions.
Experimental results show the superiority of the proposed EER approximation method in term
of computing time, and the interest of its use to reduce the learning of
parameters with genetic algorithms.
The proposed method opens new perspectives for the development of secure
multibiometrics systems with speeding up their computation time.
}
\end{abstract}


\end{frontmatter}

%\linenumbers

\section{Introduction}

Biometrics~\cite{Fons2010} is a technology allowing to recognize
people through various personal factors.
It is an active research field which design new biometric traits
from time to time
(like finger knuckle recognition~\cite{kumar-human}).
We can classify the various biometric modalities among three main
families:
\begin{itemize}
  \item
 \emph{Biological}: the recognition is based on the analysis of
  biological data linked to an individual (\emph{e.g}, DNA, EEG analysis, ...).
  \item
 \emph{Behavioural}: the recognition is based on the analysis of the behaviour of an individual
  while he is performing a specific task (\emph{e.g}, signature dynamics, gait, ...).
  \item
 \emph{Morphological}: the recognition is based on the recognition of different physical
  patterns, which are, in general, permanent and unique (\emph{e.g}, fingerprint, face
recognition, ...).
 \end{itemize}

 It is mandatory to evaluate these biometric systems in order to
 quantify their performance and compare them.

These biometric systems must be evaluated in order to compare them, or to quantify their performance.
To evaluate a biometric system, a database must be acquired (or a
common public dataset must be used).
This database must contain
as many users as possible to provide
a large number of captures of
their biometric data.
These data are separated into two different sets:
\begin{itemize}
  \item \emph{the learning} set which serves to compute the biometric reference of each user
  \item \emph{the validating} set which serves to compute their performance.
\end{itemize}
When comparing test samples to biometric references, we obtain two different kinds of scores:
\begin{itemize}
 \item \emph{the intrascores} represent comparison scores between
 the biometric reference (computed thanks to the learning set) of an individual and biometric query
 samples (contained in the validating set)
 \item \emph{the interscores} represent
comparison scores between the biometric reference of an individual
and the biometric query samples of the other individuals.
\end{itemize}
From these two sets of scores, we can compute various error rates, from which the
EER is one
functioning point which represents a very interesting error rate often used to
compare biometric systems.
In order to have reliable results, it is necessary to evaluate the performance of biometric system with huge datasets.
These huge datasets produce numbers of scores.
As the time to evaluate the performance of a biometric system depends on the
quantity of available scores, we can see that evaluation may become very long on
these large datasets.
In this paper, we present a very fast way to compute this error rate, as well as
its confidence interval in a non parametric way, on different datasets of the
literature.

Nevertheless, there will always be users for which one modality (or
method applied to this modality) will give bad results.
These low performances can be implied by different facts: the quality
of the capture, the acquisition conditions, or the individual itself.
Biometric multi-modality (or multibiometrics) allows to compensate this problem while obtaining better
biometric performances (\emph{i.e.}, better security by
accepting less impostors, and better usability
by rejecting less genuine users)
by expecting that the errors of the different modalities are not
correlated. So, the aim of multibiometrics is to protect logical or physical
access to a resource by using different biometric captures.
We can find different types of multibiometrics systems. Most of them are listed
in~\cite{ross2006handbook}, they use:
  \begin{enumerate}
    \item
 Different sensors of the same modality (\emph{i.e.},
  capacitive or resistive sensors for fingerprint acquisition);
    \item
 Different representations of the same capture
    (\emph{i.e.}, use of points of interest or texture);
   \item
 Different biometric modalities (\emph{i.e.}, face and fingerprint);
    \item
 Several instances of the same modality (\emph{i.e.}, left
    and right eye for iris recognition);
   \item
 Multiple captures (\emph{i.e.}, 25 images per second in a
   video used for face recognition);
   \item
 An hybrid system composed of the association of the
    previous ones.
\end{enumerate}

In the proposed study, we are interested in the first four kinds of
multi-modality.
We also present in this paper, a new multibiometrics approach using various fusion functions
parametrized by genetic algorithms using a fast EER (Equal Error Rate) computing method
to speed up the fitness evaluation.
%Our generated functions give better results
%than the $sum$ one which is commonly accepted as a good fusion function in the literature. The
%computation time needed to estimate the parameters of these functions is greatly
%improved by the help of our fast EER computing method.\\

This paper is related to high performance computing, because algorithms are
designed to work in an infrastructure managing the biometric authentication of
millions of individuals (\emph{i.e.}, border access control, logical acces control to webservices).
To improve the recognition rate of biometric systems, it
is necessary to regularly update the biometric reference to take into account intra class
variability. With the proposed approach, the time taken
to update the biometric reference would be lowered.
The faster is the proposed method, the more
we can launch the updating process (or the more users we can add
to the process).
We also propose an adaptation of the proposed EER computing method which gives confidence
intervals in a non parametric way (\emph{i.e.}, by computing the EER several
times through a bootstraping method).
The confidence intervals are computed on a single CPU, on several CPUs on the
same machine and on several machines.\\

The main hints of the papers are:
\begin{itemize}
 \item the proposition of a new method to approximate the EER and
 its confidence interval in a fast way
 \item the proposition of two original functions for multibiometrics fusion.
\end{itemize}

The plan is organized as following.
Section 2 presents the background of the proposed work.
Section 3 presents the proposed method for computing the approximated value
of the EER and its confidence interval.
Section 4 validates them.
Section 5 presents the proposed multibiometrics fusion functions and their performance in
term of biometric recognition and computation time against the baseline.
Section 6 gives perspectives and conclusions of this paper.

\section{Background}

\subsection{Evaluation of Biometric Systems}
\label{sec:evaluation_overview}
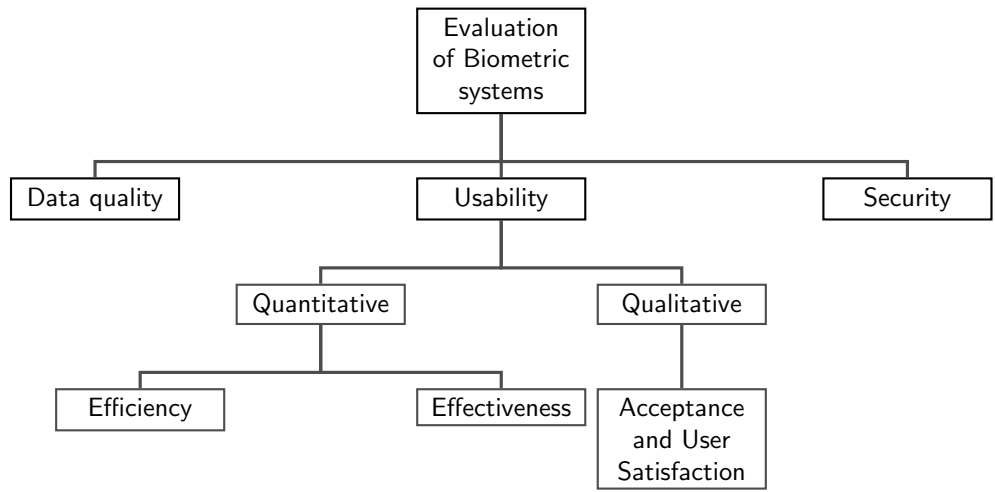
\begin{figure}[h!]
\begin{tikzpicture}[
    %Mode d'une version
    event/.style={rectangle, thick, draw, text width=2.0cm, text
centered,font=\sffamily,anchor=north},
% Children and edge
    edge from parent/.style={very thick, draw=black!70},
    edge from parent path = {(\tikzparentnode.south) -- ++(0,-1.05cm) -| (\tikzchildnode.north)},
    level 1/.style={sibling distance=9cm,level distance=1.4cm, growth parent anchor=south,nodes=event},
    level 2/.style={sibling distance=8cm},
    scale=0.6
]

\node(root) [event] {Evaluation of Biometric systems}
        child{node (Data quality) {Data quality}}
        child{node (Usability) {Usability}
                child{node (Quantitative) {Quantitative}
                           child{node (Efficiency) {Efficiency}}
                                    child{node (Effectiveness) {Effectiveness}}
                        }
                child{node (Qualitative) {Qualitative}
                                child{node (Acceptance and User Satisfaction) {Acceptance and User Satisfaction}}
                                }
        }
        child{node (Security) {Security}};
\end{tikzpicture}
\caption{Evaluation aspects of Biometric Systems.}
\label{fig:evaluation_aspects}
\end{figure}

Despite the obvious advantages of this technology in enhancing and facilitating the authentication process,
its proliferation is still not as much as attended \cite{jain2004biometrics}. As argued
in the previous section, biometric systems present several drawbacks in terms of precision,
acceptability, quality and security. Hence, evaluating biometric systems is considered as a
challenge in this research field. Nowadays, several works have been done in the literature
to evaluate such systems. Evaluating biometric systems is generally realized within three
aspects as illustrated in figure \ref{fig:evaluation_aspects}: usability, data quality and security.

\subsubsection{Usability}
According to the International Organization for Standardization
ISO 13407:1999 \citep{iso1999human}, usability
is defined as ``\textit{The extent to which a product can be used by specified users to
achieve specified goals with effectiveness, efficiency, and satisfaction in a specified context of use}''.
\begin{itemize}
    \item \emph{Efficiency} which means that users must be able to accomplish the tasks
easily and in a timely manner. It is generally measured as task time;
    \item \emph{Effectiveness} which means that users are able to complete the desired
tasks without too much effort. It is generally measured by common metrics include
completion rate and number of errors such failure-to-enroll rate (FTE) \cite{iso2006performance};
    \item \emph{User satisfaction} which measures users' acceptance and satisfaction
regarding the system. It is generally measured by studying several properties such as
easiness to use, trust in the system, \emph{etc.} The
acceptability of biometric systems is affected by several factors.
According to \cite{smith2003human}, some members of the human-computer interaction (HCI)
community believe that interfaces of security systems do not reflect good thinking in terms
of creating a system that is easy to use, while maintaining an acceptable level of security.
Existing works \cite{coventry2003usability, gunson2011usability} show also that there is a
potential concern about the misuse of personal data (\emph{i.e.}, templates) which is
seen as violating users' privacy and civil liberties.
Moreover, one of our previous work
\cite{elabed2010usability} shows the necessity of taking into account users' acceptance and
satisfaction when designing and evaluating biometric systems. More generally speaking,
even if the performance of a biometric system outperformed another one, this will
not necessarily mean that it will be more operational or acceptable;\\
\end{itemize}

\subsubsection{Data quality}
It measures the quality of the biometric raw data
\cite{tabassi2005quality, elabed2010quality}. Low quality samples increase the enrollment
failure rate, and decrease system performance.
Therefore, quality assessment is considered as a crucial factor required in both the
enrollment and verification phases. Using quality information, the bad quality samples can
be removed during enrollment or rejected during verification. Such information could also
be used in soft biometrics or multimodal approaches \cite{Krzysztof2009quality}. Such type of
assessment is generally used to quantify biometric sensors, and could be also used to
enhance system performance;\\

\subsubsection{Security}
It measures the robustness of a biometric system (algorithms, architectures and devices)
against attacks. Many works in the literature
\citep{ratha2001security, uludag2004attacks, galbally2010attacks}
show the vulnerabilities of biometric systems which can considerably decrease their security. Hence,
the evaluation of biometric systems in terms of security is considered as an important factor to ensure
its functionality. The International Organization for Standardization
ISO/IEC FCD 19792 \cite{iso2008security} addresses the
aspects of security evaluation of such systems. The report presents an overview of
biometric systems vulnerabilities and provide some recommendations to be taking into account during
the evaluation process. Nowadays, only few partial security analysis studies with relation to biometric
authentication systems exist. According to ISO/IEC FCD 19792 \cite{iso2008security}, the security
evaluation of biometric systems is generally divided into two complementary assessments:
1) assessment of the biometric system (devices and algorithms) and
2) assessment of the environmental (for example, is the system is used indoor or outdoor?)
and operational conditions (for example, tasks done by system administrators to ensure that
the claimed identities during enrolment of the users are valid). A type-1 security assessment method
is presented in a personal previous work \cite{elabed2011security}. The proposed method has shown its efficiency
in evaluating and comparing biometric systems.

\subsection{Performance Evaluation of Biometric Systems}
\label{sec:performance_evaluation}
The performance evaluation of biometric systems is now carefully considered in
biometric research area. We need a reliable evaluation methodology in order to put
into obviousness the benefit of a new biometric
system. Nowadays, many efforts have been done to achieve this
objective. We present in section \ref{sec:performance_evaluation} an overview of the
performance metrics, followed by the research benchmarks in biometrics as
an illustration of the evaluation methodologies used in the literature
for the comparison of biometric systems.

\subsubsection{Performance metrics}
\label{sec:performance_evaluation}
\begin{figure}[!ht]
\centering
\includegraphics[width=0.7\linewidth]{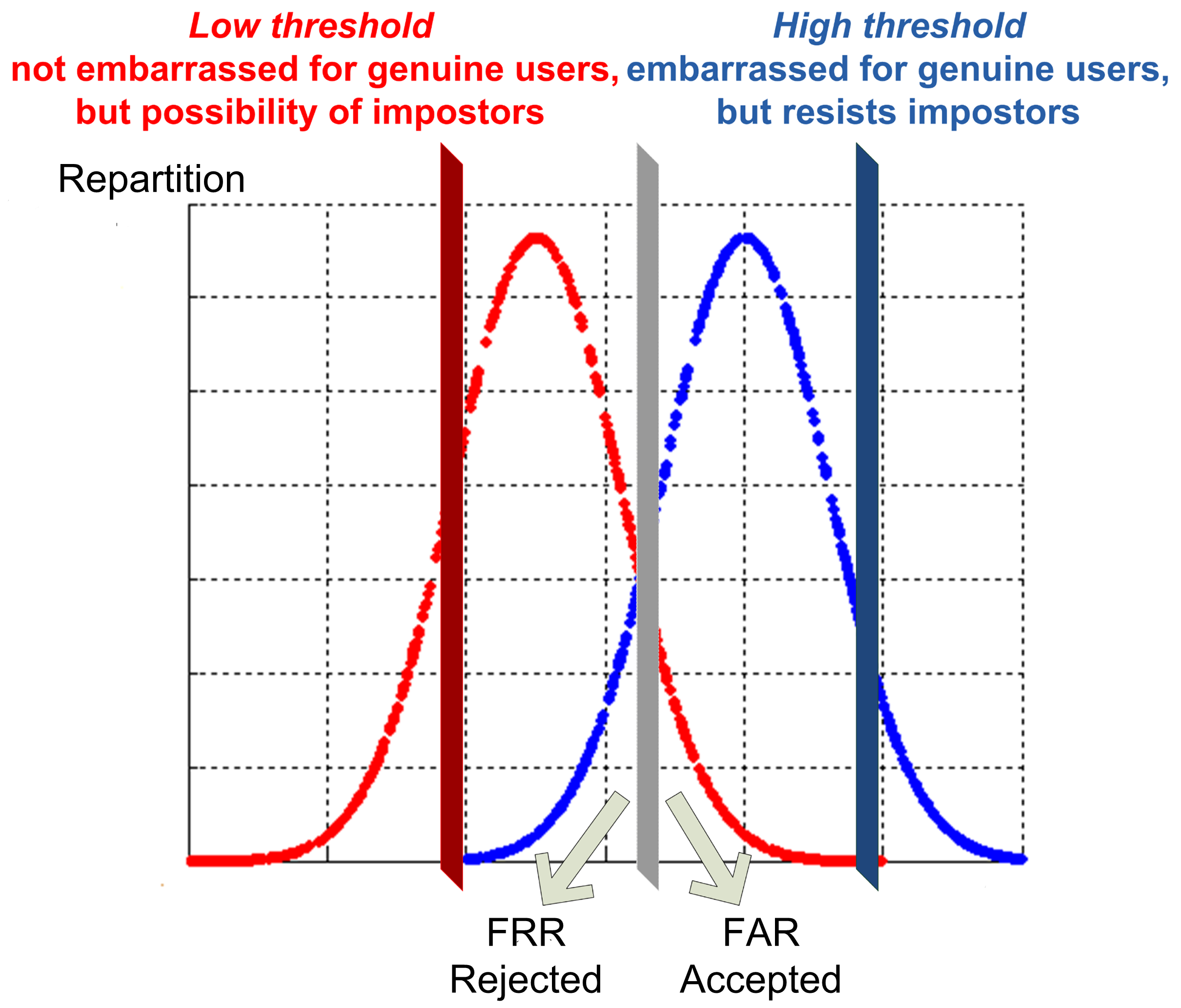}
\caption{Distribution of genuine users and impostor scores.}
\label{fig:scores_distribution}
\end{figure}

By contrast to traditional methods, biometric systems do not provide a cent per cent reliable
answer, and it is quite impossible to obtain such a response. The comparison result between
the acquired biometric sample and its corresponding stored template is illustrated by a
distance score. If the score is lower than the predefined decision threshold,
then the system accepts the claimant, otherwise he is rejected. This threshold
is defined according to the security level required by the application.
Figure \ref{fig:scores_distribution} illustrates the theoretical distribution of the
genuine and impostor scores. This figure shows that errors depend from the used threshold. Hence,
it is important to quantify the performance of biometric systems. The International
Organization for Standardization ISO/IEC 19795-1 \cite{iso2006performance} proposes several
statistical metrics to characterize the performance of a biometric system such as:

\begin{itemize}
    \item \emph{Failure-to-enroll rate (FTE)}: proportion of the user population for whom the
biometric system fails to capture or extract usable information from biometric sample;
    \item \emph{Failure-to-acquire rate (FTA)}: proportion of verification or identification
attempts for which a biometric system is unable to capture a sample or locate an
image or signal of sufficient quality;
    \item \emph{False Acceptation Rate (FAR)} and \emph{False Rejection Rate (FRR)}: FAR is the proportion
of impostors that are accepted by the biometric system, while the FRR is the proportion
of authentic users that are incorrectly denied. The computation of these error rates
is based on the comparison of the scores against a threshold (the direction of the
comparison is reversed if the scores represent similarities instead of distances).
FRR and FAR are respectively computed (in the case of a distance score) as in (\ref{eq:frr})
and (\ref{eq:far}), where $intra_i$ (respectively $inter_i$) means the intra score at
position $i$ in the set of intra score (respectively inter score at position
$i$) and $Card(set)$ is the cardinal of the set in argument, $thr$ is the decision
threshold, and $\mathds{1}$ is the indicator function.

\begin{equation}
  \label{eq:frr}
  FRR = \frac{\sum_{score \in intra}\mathds{1}\{score > thr\}}{Card(intra)}
\end{equation}

\begin{equation}
  \label{eq:far}
  FAR = \frac{\sum_{score \in inter}\mathds{1}\{score \le thr\}}{Card(intra)}
\end{equation}

\item \emph{Receiver operating characteristic (ROC) curve}: the ROC curve is obtained by
computing the couple of (FAR, FRR) for each tested threshold. It plots the FRR versus the FAR.
The aim of this curve is to present the tradeoff between FAR and FRR and to have a quick
overview of the system performance and security for all the parameters configurations.

    \item \emph{Equal error rate (EER)}: it is the value where both errors rates,
FAR and FRR, are equals (\emph{i.e.}, FAR = FRR). It constitutes a good indicator,
and the most used, to evaluate and compare biometric systems.
In other words, lower the EER value is,
higher the accuracy of the system. Using the ROC curve, the EER is computed by selecting the
couple of (FAR, FRR) having the smallest absolute difference (\ref{eq:precision}) at the
given threshold $\tau$:

\begin{equation}
  \label{eq:precision}
  \tau = \underset{\tau}{\operatorname{argmin}}
               (abs(FAR_{\tau}-FRR_{\tau})),
            \forall_{\tau \subset Card\{ROC\}} \nonumber
\end{equation}

\noindent
 and returning their average
    (\ref{eq:eer}):

\begin{equation}
  \label{eq:eer}
  EER = \frac{FAR_{\tau} + FRR_{\tau}}{2}
\end{equation}

\noindent By this way, we have obtained the best approaching EER value
with the smallest precision error.
The classical EER computing algorithm is presented in the Figure~\ref{fig:sloweer}
\footnote{another, slower, way of computing would be to
test each unique score of the intrascores and interscores sets, but this would
held a too important number of iterations. We named it ``whole'' later in the
paper.}.
From Figure~\ref{fig:sloweer}, we can see that
the complexity is in $O(n*m)$ with $n$ the
number of thresholds held in the computation, and, $m$ the number of scores in
the dataset. As it is impossible to reduce $m$, we have to find a better method
which reduces $n$. We did not find, in the literature, methods allowing to reduce
computation time in order to obtain this EER.

\begin{figure}[h!]
\footnotesize
\begin{algorithmic}
\STATE $ROC \leftarrow [] $
\STATE $EER \leftarrow  1.0 $
\STATE $DIFF \leftarrow 1.0 $
\STATE $START \leftarrow min(scores)$
\STATE $END \leftarrow max(scores)$

\FOR{$\tau$ $START$ to $END$ in $N$ steps}
  \STATE  $FAR \leftarrow$ compute $FAR$ for $\tau$
  \STATE  $FRR \leftarrow$ compute $FRR$ for $\tau$
    \STATE  append $(FAR,FRR)$ to $ROC$
    \IF{$abs(FAR-FRR)<DIFF$}
    \STATE $DIFF \leftarrow abs(FAR-FRR)$
        \STATE $EER \leftarrow (FAR+FRR)/2$
    \ENDIF
\ENDFOR

\RETURN $EER$, $ROC$
\end{algorithmic}
\caption{Classical EER computing algorithm.}
\label{fig:sloweer}
\end{figure}

\end{itemize}

\subsubsection{Biometrics Datasets}
\label{sec:biometrics_benchmarks}
A public dataset allows researchers to test their algorithm
and compare them with those from the state of the art. It
takes a lot of time and energy to build a large and significant
dataset. It is very convenient to download one for research
purposes. We present in this section an overview of the used
datasets in this paper. Table \ref{tab:databases} presents a summary of
these datasets.

\begin{itemize}

    \item \textbf{Biometric Scores Set - Release 1 (BSSR1)}\\
The BSSR1~\cite{bssr1url} database is an ensemble of scores sets from different
biometric systems. In this study, we are interested in the subset containing the
scores of two facial recognition systems and the two scores of a fingerprint
recognition system applied to two different fingers for $512$ users.
This database has been used many times in the
literature~\cite{nandakumar2008likelihood,sedgwick2005preliminary}.

    \item \textbf{BANCA}\\
The second database is a subset of scores produced from the BANCA
database~\cite{bancascores}. The selected scores correspond to the following one labelled:
\begin{enumerate}
 \item IDIAP\_voice\_gmm\_auto\_scale\_25\_100\_pca.scores
 \item SURREY\_face\_nc\_man\_scale\_100.scores
 \item SURREY\_face\_svm\_man\_scale\_0.13.scores
 \item UC3M\_voice\_gmm\_auto\_scale\_10\_100.scores
\end{enumerate}
%We have chosen empirically this subset, not on the performance of the methods.

The database as two subsets G1 and G2.
G1 set is used as the learning set, while G2 set is used as the validation
set.

    \item \textbf{PRIVATE}\\
The last database is a chimeric one we have created for this purpose by
combining two public biometric template databases: the AR~\cite{martinez1998ar}
for the facial recognition and the GREYC keystroke~\cite{giot2009database} for
keystroke dynamics~\cite{Monrose2000351,giot2011keystroke}.
The AR database is composed of frontal facial images of $126$ individuals under
different facial expressions, illumination conditions or occlusions. These images
have been taken during two different sessions with $13$ captures per session.
The GREYC keystroke contains the captures on several sessions on two months of
$133$ individuals.
Users were asked to type the password "greyc laboratory" $6$
times on a laptop and $6$ times on an USB keyboard by interlacing the typings (one time on a keyboard, one time on another).
We have selected the first $100$ individual of the AR database and we have
associated each of these individuals to another one in a subset of the GREYC
keystroke database having $5$ sessions of captures. We then used the $10$ first
captures to create the biometric reference of each user and the $16$ others to compute the
intra and inter scores.
These scores have been computed by using two different methods for the face
recognition and two other ones for the keystroke dynamics.
\end{itemize}

\begin{table}[!ht]
  \caption{Summary of the different databases used to validate the proposed method}
  \small
  \label{tab:databases}
  \centering
  \begin{tabular}{|l|r|r|r|}\hline
    \textbf{Nb of} & \textbf{BSSR1} & \textbf{PRIVATE} & \textbf{BANCA}\\\hline\hline
    users & 512 & 100 & 208 \\ \hline
    intra tuples & 512 & 1600 & 467 \\ \hline
    inter tuples & 261632 & 158400 & 624 \\ \hline
    items/tuples  & 4 & 5 & 4 \\ \hline
\end{tabular}
\end{table}

\subsection{Multibiometrics}
We focus in this part on the state of the art on multimodal systems involving biometric modalities usable for all computers (keystroke, face, voice...).
The scores fusion is the main process in multimodal systems. It can be operated on the scores provided by algorithms or in the templates
themselves~\cite{raghavendra2009pva}.
In the first case, it is necessary to normalize the different scores as they may not evolve in the same range.
Different methods can be used for doing this, and the most efficient methods are
\emph{zscore}, \emph{tanh} and \emph{minmax}~\cite{jain2005snm}.
Different kinds of fusion methods have been applied on biometric systems. The fusion can
be done with multiple algorithms of the same modality.
For example, in~\cite{hocquet2007}, three different keystroke dynamics
implementations are fused with an improvement of the EER, but
less than 40 users are involved in the database. In~\cite{teh2007statistical},
two keystroke dynamics systems are fused together by using
weighted sums for 50 users, but no information on the weight computing is provided.
The fusion can also be done within different modalities in order to
improve the authentication process. In~\cite{roli2008adaptive}, authors
use both face and fingerprint recognition, the impact of error rate reduction is
used to reduce the error when adapting the user's biometric reference.
There is only one paper (to our knowledge) on keystroke dynamics fusion with
another kind of biometric modality (voice recognition): it is presented
in~\cite{montalvao2006mbf}, but only 10 users are involved in the experiment.
In~\cite{fierrez2003comparative}, multi-modality is done on fingerprints, speech,
and face images on 50 individuals. Fusion has been done with
SVM~\cite{vapnik1996theory} with good improvements, especially,
when using user specific classifiers.

Very few multimodal
 systems have been proposed for being used in classical computers and the published ones have been validated on small databases.
In order to contribute to solve this problem, we propose a new approach in the
following section.

\section{Fast EER and Confidence Intervals Estimation}
We propose a kind of dichotomic EER computing function, in order to quickly approximate its value.
Thanks to this computing speed up, we can use it in time consuming applications.
Finally, we present a confidence interval computing method based on our
approximated EER calculation associated to parallel and distributed computing.

\subsection{EER Estimation}

Computation time to get the EER can be quite important. When the EER value needs
to be computed a lot of time, it is necessary to use a faster way than the standard one.
In the biometric community, the shape of the ROC curve always follows the same pattern:
it is a monotonically decreasing function (when we present FRR against FAR, or
increasing when we present 1-FRR against FAR) and the EER value is the curve's point
having $x_{ROC}=y_{ROC}$ (or $FAR=FRR$). Thanks to this fact, the curve symbolising the
difference of $y_{ROC}$ against $x_{ROC}$ is also a monotonically decreasing function from
$1$ to $-1$, where the point at $y_{DIFF}=0$ represents the EER (and its
value is $x_{DIFF}$ because $x_{ROC}=y_{ROC}$ or $FAR=FRR$).
With these information, we know that to get the EERs, we need to find the
$x_{DIFF}$ for which $y_{DIFF}$ is the closest as possible to zero. An analogy with
the classical EER computing, would be to incrementally compute $y_{DIFF}$ for
each threshold by increasing order and stop when $y_{DIFF}$ changes of sign. By this
way, we can expect to do half thresholds comparisons than with the classical
way if scores are correctly distributed.
A clever way is to use something equivalent to a divide and conquer algorithm
like the binary search and obtain a mean complexity closer to $O(log(n))$.
That is why we have implemented a polytomous version of EER computing:
\begin{enumerate}
\item We chose $i$ thresholds linearly distributed on the scores sets
\item For each threshold $t$ among the $i$ thresholds, we compute the FAR and FRR values ($FAR_t$, $FRR_t$)
\item We take the two following thresholds $t1$ and $t2$ having
$sign(FRR_{t1}-FAR_{t1})$ different of $sign(FRR_{t2}-FAR_{t2})$
\item We repeat
step 2 with selecting $i$ thresholds between $t1$ and $t2$ included while
$FRR_{t1}-FAR_{t1})$ does not reach the attended precision.
\end{enumerate}

By this way, the number of threshold comparisons is far smaller than
in the classical way. Its complexity analysis is not an easy task
because it depends both on the attended precision and the choice of
$i$. It can be estimated as O($\log(N)$).

\begin{figure}[!htb]
\subfloat[Iteration 1]{\includegraphics[width=0.5\linewidth]{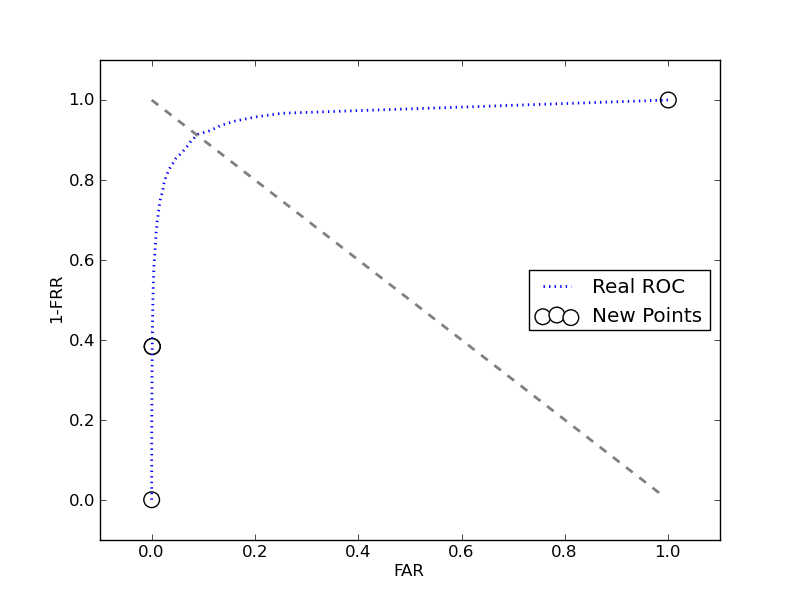}}
\subfloat[Iteration 2]{\includegraphics[width=0.5\linewidth]{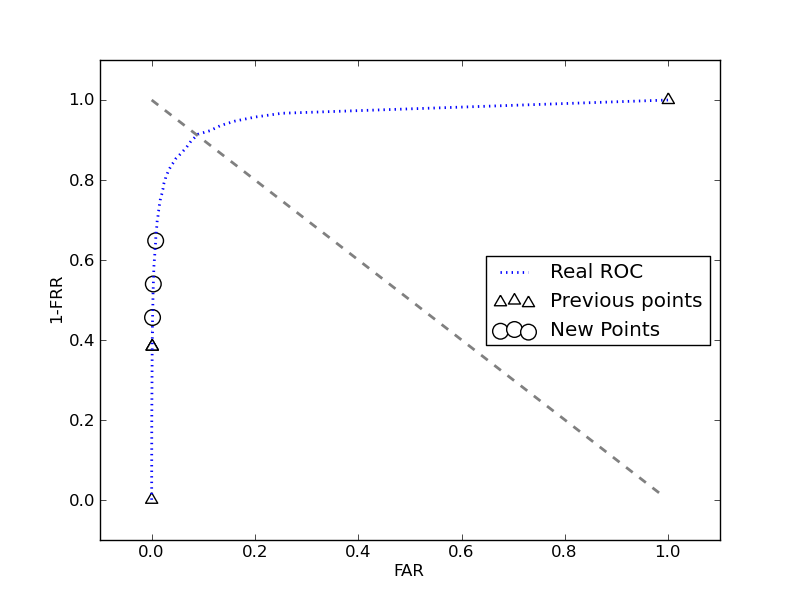}}\\
\subfloat[Iteration 3]{\includegraphics[width=0.5\linewidth]{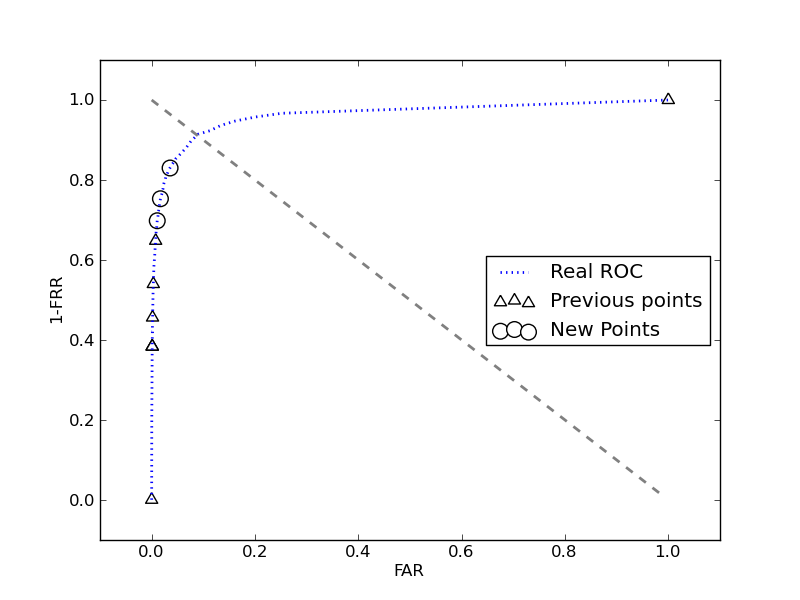}}
\subfloat[Iteration 4]{\includegraphics[width=0.5\linewidth]{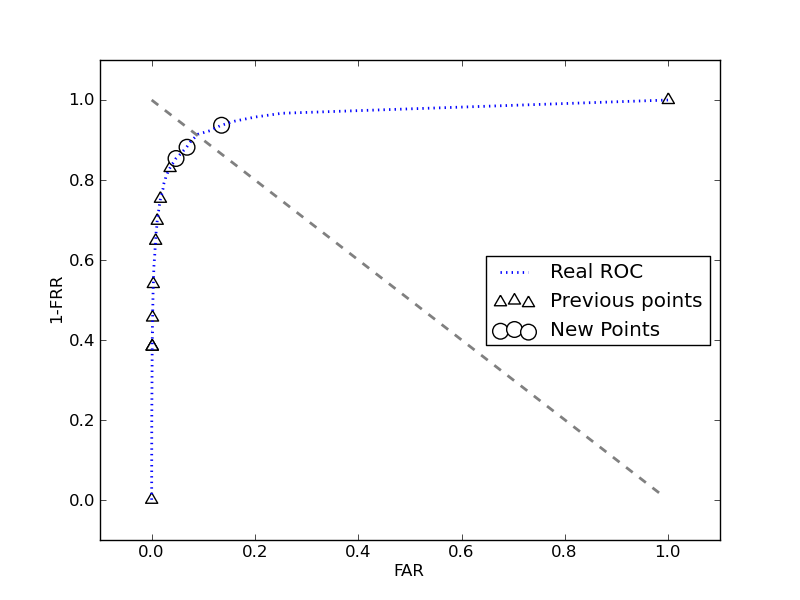}}\\
\subfloat[Iteration 5]{\includegraphics[width=0.5\linewidth]{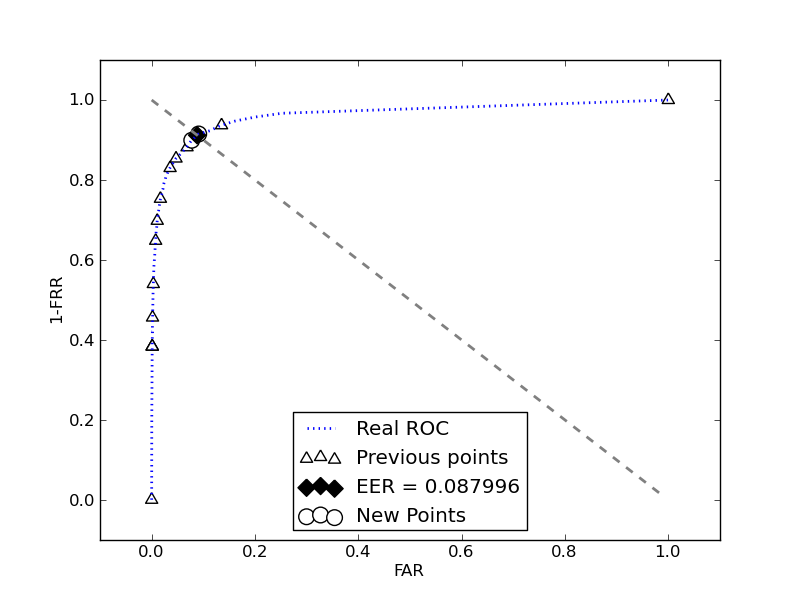}}
\subfloat[ROC curve]{\label{fig:expleer_roc}\includegraphics[width=0.45\linewidth]{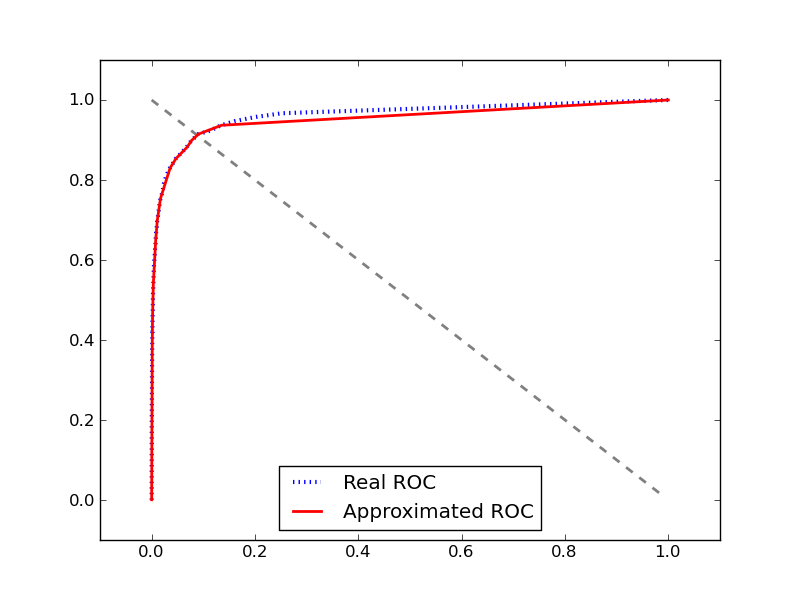}}
    \caption{Points computed by the proposed algorithm when $i=5$.
In this case, the EER value is found in five iterations.
Each image represents an iteration, with: the real ROC curve, the points computed at the iteration and the points computed at the previous iteration (different thresholds may produce the same points).}
    \label{fig:expleer}
\end{figure}

Figure~\ref{fig:fasteer} presents the algorithm while Figure~\ref{fig:expleer}
illustrates it by showing the different iterations until getting the EER value
with a real world dataset.
We have chosen $i=5$ points to compute during each iteration.
The EER is obtained in five iterations.
Circle symbols present the points computed at the actual iteration,
triangle symbols present the points computed at the previous iterations,
and the dotted curve presents the ROC curve if all the points are computed.
Very few points are computed to obtained the EER value.
Figure~\ref{fig:expleer_roc} presents the real ROC curve and the ROC curve
obtained with the proposed method.
We can see that even if we obtain an approximated version of the real ROC curve,
it is really similar around the EER value (cross with the dotted lined).

\begin{figure}[!htb]
\footnotesize
\begin{algorithmic}
\STATE $ROC \leftarrow [] $
\STATE $CACHE \leftarrow \{\} $
\STATE $START \leftarrow min(scores)$
\STATE $END \leftarrow max(scores)$

\WHILE{True}
    \FOR{$THRESHOLD$ from $START$ to $END$ in $N$ steps}
        \STATE $SDIFF \leftarrow []$
        \STATE $THRESHOLDS \leftarrow []$
        \IF{not empty $CACHE[THRESHOLD]$}
            \STATE $FAR,FRR \leftarrow CACHE[THRESHOLD]$
        \ELSE
            \STATE  $FAR \leftarrow$ compute $FAR$ for $THRESHOLD$
            \STATE  $FRR \leftarrow$ compute $FRR$ for $THRESHOLD$
            \STATE  append $(FAR,FRR)$ to $ROC$
            \STATE $CACHE[THRESHOLD] \leftarrow (FAR, FRR)$
        \ENDIF

         \IF{ $abs(FAR-FRR) < PRECISION$ }
        \STATE $EER  \leftarrow (FAR+FRR)/2$
        \RETURN $EER$, $ROC$
        \ENDIF

        \STATE append $FAR-FRR$ to $SDIFF$
        \STATE append $THRESHOLD$ to $THRESHOLDS$
    \ENDFOR

    \STATE $PSTART \leftarrow -1$
    \STATE $PEND \leftarrow -1$
  \FOR{ $PIVOT=0$ to $STEPS-1$}
      \IF{ $sign(SDIFF[PIVOT]) \ne sign(SDIFF[PIVOT+1])$}
                \STATE $PSTART \leftarrow PIVOT$
                \STATE $PEND \leftarrow PIVOT+1$
        \STATE \textbf{break}
            \ENDIF
    \ENDFOR
    \COMMENT{PSTART and PEND are set}

    \STATE $START \leftarrow THRESHOLDS[PSTART]$
    \STATE $END \leftarrow THRESHOLDS[PEND]$
\ENDWHILE

\end{algorithmic}
\caption{Fast EER Computing Algorithm}
\label{fig:fasteer}
\end{figure}

\clearpage
\subsection{Confidence Intervals Estimation}
We also provide a method to compute the confidence interval of the EER value.
It is based on a bootstrapping method and can be used in a parallelized way.

\subsubsection{Bootstrapping}
It is interesting to give a confidence interval of an EER value, because we are not
totally sure of its value.
One way is to obtain this confidence interval parametrically, but it requires to
have strong hypothesis on the function of the EER value
(the scores come
from independant and identically distributed variables, even for the scores of
the same user).
As such assumption is too strict (and probably false), it is possible to use non parametric
confidence intervals.
One of these non parametric methods is called
``bootstrap''~\cite{johnson2001introduction}.
Such method is often used when the distribution is unknown or the number of
samples is too low to correctly estimates the interval.
The main aim is to re-sample the scores several times, and, compute the EER value for
each of these re-sampling.
The boostraping method works as following:
\begin{enumerate}
  \item Use the $intra$ and $inter$ scores to compute the EER $\hat{\chi}$.
  \item Resample $K$ times the $intra$ and $inter$ scores and store them in
$intra^i$ and $inter^i$ ($0<i<=K$).
  \begin{itemize}
    \item Generate the resampled $intra^i$ scores by sampling $Card(intra)$
scores with replacement from $intra$.
    \item Generate the $inter^i$ scores by sampling $Card(inter)$
scores with replacement from $inter$.
  \end{itemize}
  \item Compute the $K$ EERs ($\chi^i$)  for each couple of $intra^i$ and
$inter^i$ ($0<i<=K$).
  \item Store the $K$ residuals $e^i = \hat{\chi} - \chi^i$.
  \item The $100(1-\alpha)$\% confidence interval of the EER is formed by taking the
interval from $\hat{\chi}-e_{Upper}$ to $\hat{\chi}-e_{Lower}$, with $e_{Lower}$
and $e_{Upper}$ which
respectively represent the $\alpha/2$th and the $1-\alpha/2$th percentiles of
the distribution of $e$.
\end{enumerate}

\begin{figure}[!htb]
\includegraphics[width=.9\linewidth]{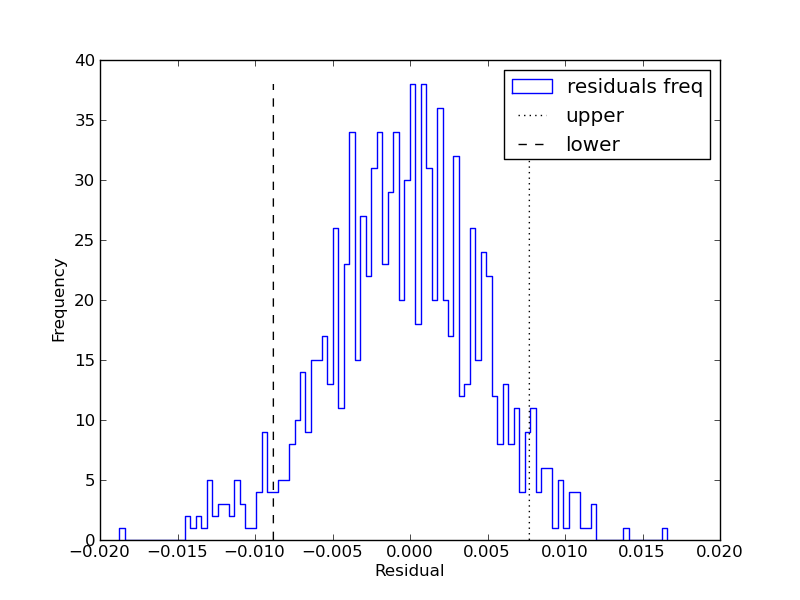}
\caption{Histogram of residuals and $\alpha/2$ and $1-\alpha/2$ percentiles for
a confidence interval at 90\%.}
\label{fig:residuals}
\end{figure}

Figure~\ref{fig:residuals} presents the residuals of one run of the bootstraped
method on
a real world dataset with the lower and upper limits used to compute the confidence
interval.
We can see that the consuming part of this algorithm is the fact to compute $K$
times the EER.
As all the EER computing is totally independent from each other, the $K$
computations can be done in a parallel way.

\subsubsection{Parallelization}
\begin{figure}[!htb]
\centering
\subfloat[Single]{\includegraphics[width=.35\linewidth]{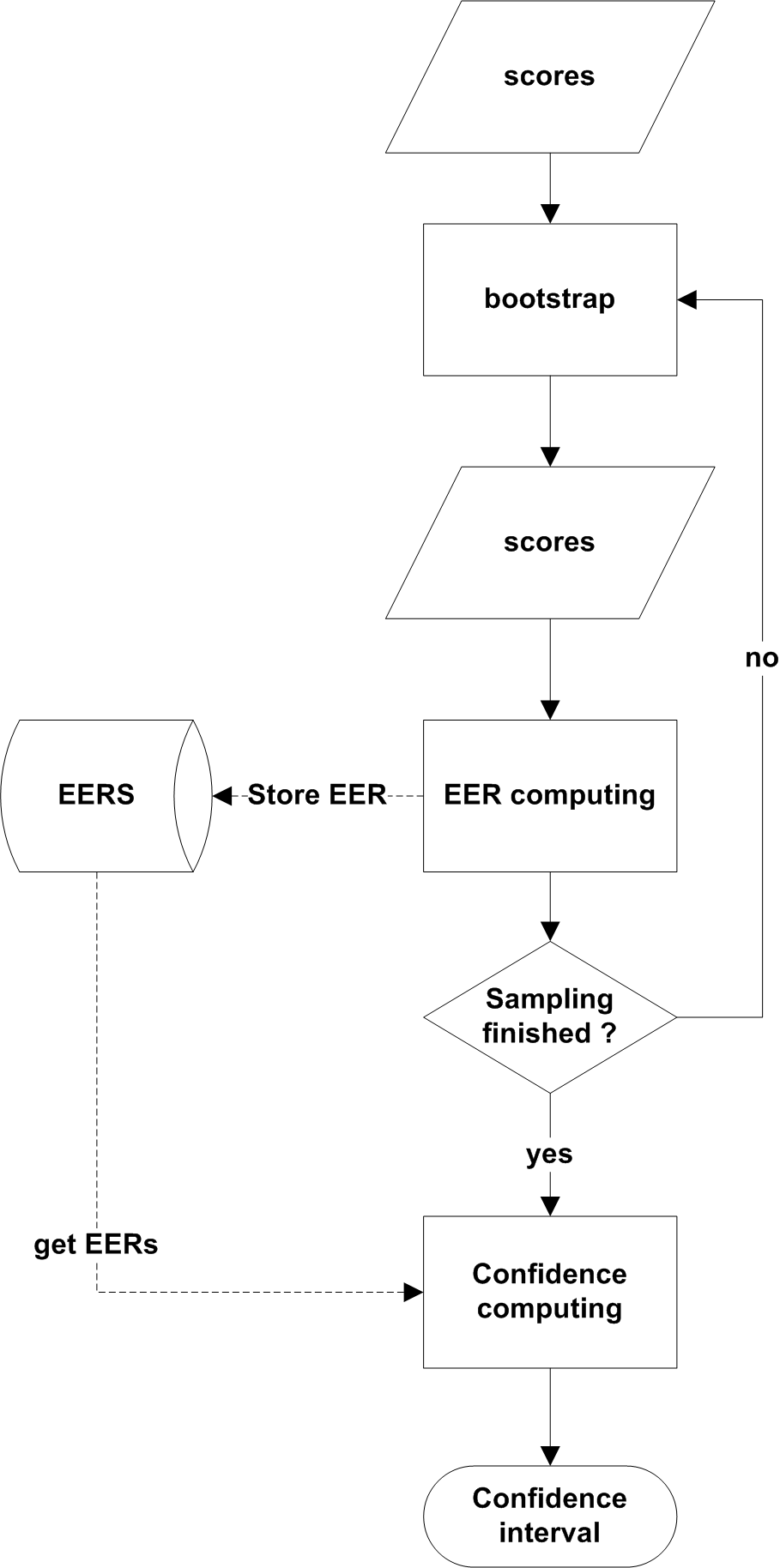}\label{fig_archi_conf_single}}
\subfloat[Parallel]{\includegraphics[width=.55\linewidth]{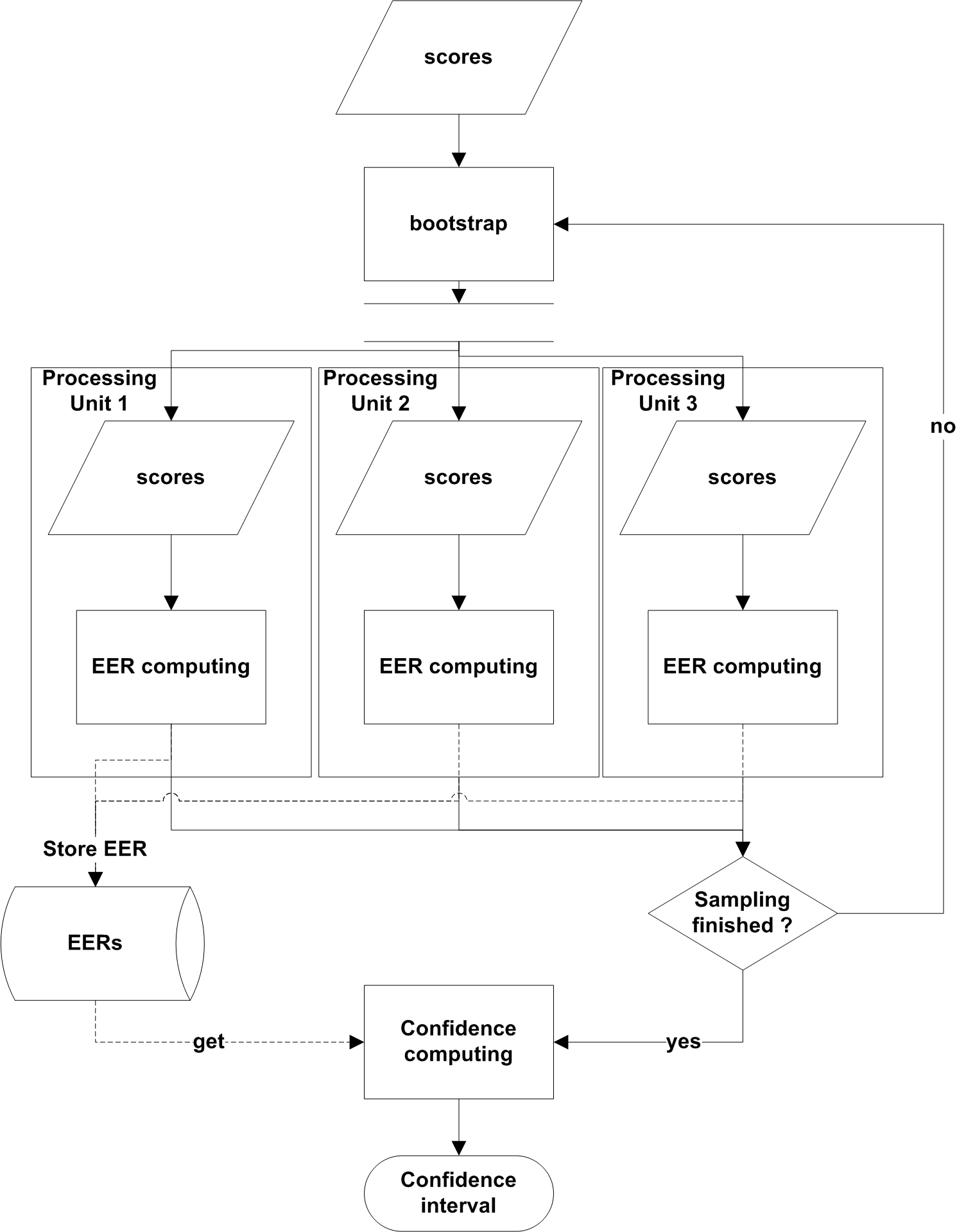}\label{fig_archi_conf_parallel}}\\
\subfloat[Distributed]{\includegraphics[width=.55\linewidth]{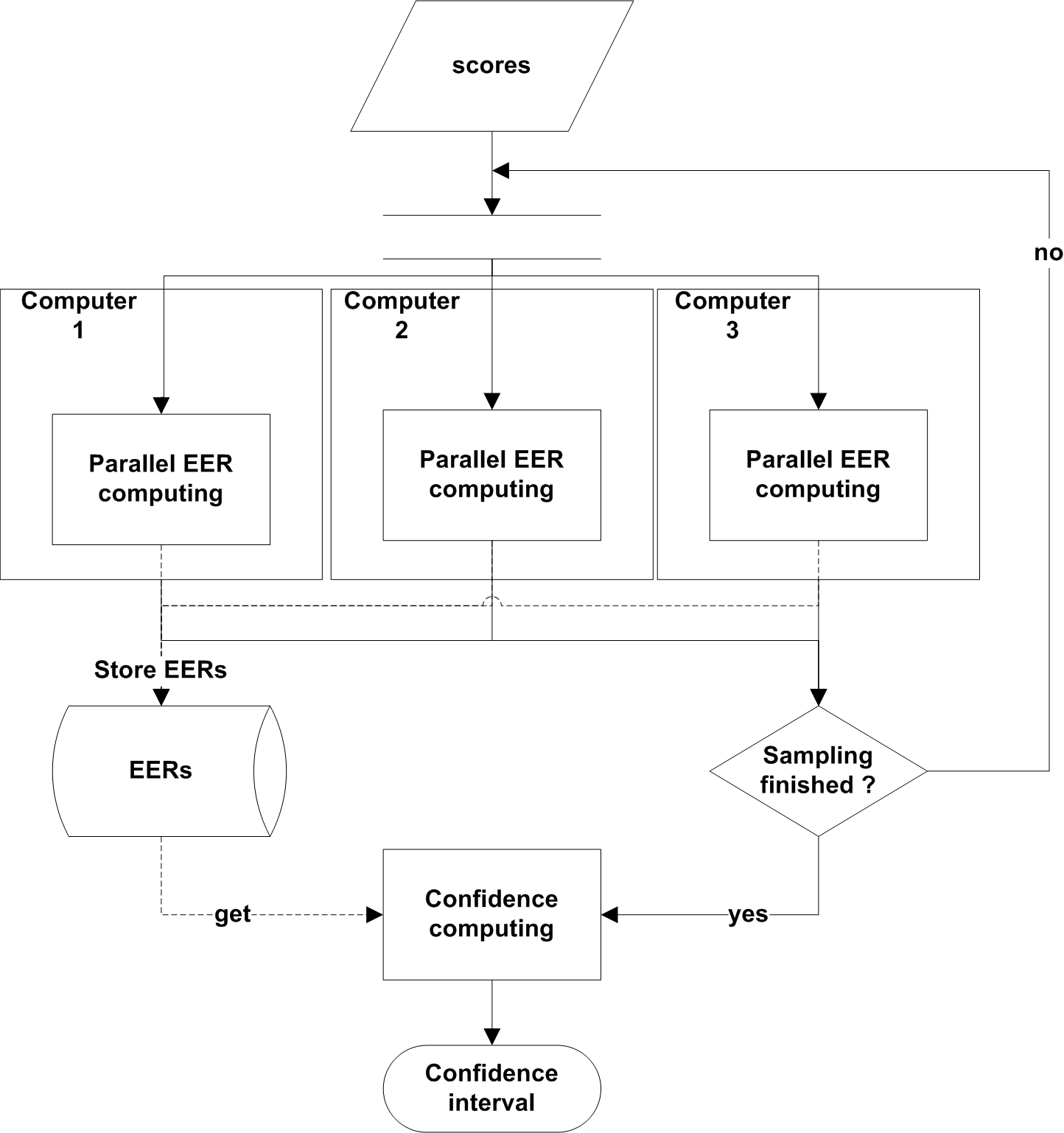}\label{fig_archi_conf_distributed}}
\caption{Different architectures to compute the confidence
interval.\label{fig_archi_conf}}
\end{figure}

We propose three different ways to compute the confidence interval:
\begin{itemize}
\item \emph{Single}.
The single version consists in doing all the computations
in a sequential manner (\emph{cf.} figure~\ref{fig_archi_conf_single}).
The EER with the original scores is computed. The whole
set of resampled scores is created.
The EER of each new distribution is
computed.
The confidence interval is computed from the results.

\item \emph{Parallel}.
The parallel version consists in using the several cores or processors on the
computer used for the computation (\emph{cf.}
figure~\ref{fig_archi_conf_parallel}).
In this case, several EERs may be computed at the same time (in the better case,
with a computer having $n$ processing units, we can compute $n$ different
results at the same time).
The procedure is the following: the EER with the original scores is computed.
The whole set of resampled scores is created.
The main program distributes the $K$ EERs computations on the different processing
units.
Each processing unit computes an EER and returns the results, until the main
program stops to send it new data.
The results are merged together.
The confidence interval is computed from the results.

\item \emph{Distributed}.
The distributed version consists in using several computers to improve the
computation (\emph{cf.} figure~\ref{fig_archi_conf_distributed}).
The computation is done in a parallelized way on each computer.
In this case, much more EERs can be computed at the same time.
The main program generates a set $S=\{S_1,...S_T\}$ of $T$ values symbolising
$T$ subworks, were the subwork $T_i$ must compute $S_i$ EERs (thus $\sum S=K$).
The procedure is the following: the EER with the original scores is computed.
The main program sends the intra and inter scores on each worker (\emph{i.e.}, a
computer).
The main program distributes the $T$ numbers to each worker.
Each time a work receive such number ($S_i$), it computes the $S_i$ resamples
sets.
Then, it computes (using the Parallellized way) the $S_i$ EERs by distributing
them on its processing units.
It merges the $S_i$ EERs together and send them to the main program.
The main program merges all the results together.
The confidence interval is computed from the results.
\end{itemize}

We can see that the three schemes are totally different.
The parallelized version may be used on all recent computers which have several
processing units, while the distributed version needs to use several computers connected through a network.

\clearpage

\section{Protocol}
\subsection{Databases Sets}
In order to do these evaluations, we have used three different biometric
databases presented in section~\ref{sec:biometrics_benchmarks}.

\subsection{Evaluation of the EER Computing}
The two different algorithms for EER computing have been run on five different sets of
scores (three of keystroke dynamics and two of face recognition, generated with the PRIVATE database) with various parameters.
We call \emph{classic} the classical way of computing the EER and \emph{polyto}
the proposed version of the algorithm.
The classic way is tested by using 50, 100, 500 and 1000 steps to compute the
EER. The polytomous way is tested by using between 3 and 7 steps and a precision of
0.01, 0.005 and 0.003.
The aim of these tests is to compare how the proposed method performs better than the classical
one, and what are its best parameters.\\

\subsection{Utility for Non-parametric Confidence Interval}
Confidence intervals are also an interesting information on the performance of a
system.
The properties of the score distribution may forbid the use of parametric
methods to compute it.
This is where the boostrap method helps us by computing several times the EER with
resampling methods.
We have tested the computation time of confidence interval computing for two
systems of each database, which give us six different systems.
We have computed the EER value under three different ways: the polytomous one, the
classical one (with 1000 steps), and another we called whole.
The whole method is similar to the classical one, except that it uses all the possible
scores present in the intra inter scores arrays as thresholds, instead of
artificially generating them in a predefined interval (thus, there may be far
more thresholds than in the classic method, or far less, depending on the number
of available scores in the database).
The results are then validated with confidence intervals.
The test scripts were written with the Python language.
The EER computing methods and the resampling methods have been compiled in
machine code thanks to the Cython~\cite{seljebotn2009fast} framework (which speed up computation time).
The parallelization is done with the joblib~\cite{joblib} library.
The distributed version is done by using the task management provided by
Ipython~\cite{perez2007ipython}.
The standard and parallelized versions have been launched on a recent Linux
machine with an Intel® Core™ i5 with 4 cores at 3.20GHz and 4Gb of RAM.
Four processes are launched at the same time.
For the distributed machine, the orchestrator is the same machine.
It distributes the jobs on three multicores machines (itself, another similar machine and an another
machine with an Intel® Xeon® with 8 cores at 2.27GHz and 4Gb of RAM. The
controller sends two more jobs last machine machine).

\subsection{Experimental Results}
\subsubsection{Evaluation of the EER Computing}
Table~\ref{tab:resspeed1} presents the results obtained within the first
tested biometric system. We present the name of the method, the error of
precision while computing the EER, the computation time in milliseconds and the
number of comparisons involved (each comparison corresponds to the comparison of
a threshold against the whole set of intra and inter scores).
The real computation time taken by a comparison is given
in~(\ref{eq:time}), where $n$ is the number of thresholds to compare, $A$
is the timing to do a comparison and $B$ and $C$ depends on the algorithm.
\begin{equation}
\label{eq:time}
T = n * (A*(Card\{intra\} + Card\{inter\}) + B) + C
\end{equation}
We can see that the computation time is highly related to the number of
comparisons and the size of the score set. Using the Kruskal-Wallis test at a confidence
degree equals to $95\%$, the proposed method significantly outperformed the classical method
in terms of errors (with a \emph{p} value $=0.0305$) and computation time (with a \emph{p} value $=0.002562$).
The obtained results are slightly similar for the five tested biometrics modalities. We
can observe that, in the classic method, using 50 steps gives
not enough  precise results, while using 1000 gives a very good precision, but
is really time consuming;
depending on the dataset, 500 steps seems to be a good
compromise between precision error and computation time.
In all the polytomous configurations, the computation time is far better
than the fastest classic method (50 steps) while having a greatest precision. This
precision is always better than the classic method with 100 steps and approach
or is better than the precision in 1000 steps.
This gain of time is due to the
lowest number of involved comparisons. In an $n$ steps classical computing, we
need to check $n$ thresholds, while in the polytomous way this number depends
both on the dataset and the required precision: with our dataset, it can vary
from 8 to 35 which is always lower than 50.
As the computation time depends only on
this value, we can say that the fastest algorithms are the one having the
smallest number of tests to complete. \\

Based on the number of comparisons (and the timing
computation), Table~\ref{tab:resspeedtot} presents the best results for each
modality (when several methods return the same number of iterations, the
most precise is chosen).

\begin{table}[!tb]
\centering
  \small
  \caption{
Comparison of the Different EER Computing Methods And Configurations On The
First Test Set.
LABEL presents the used method.
ERROR is the difference between FAR and FRR values.
TIME is the time needed to compute the EER value.
COMP. is the number of threshold comparison done.}
  \label{tab:resspeed1}

  \begin{tabular}{|l|c|c|c|c|}\hline
  \textbf{LABEL} & \textbf{ERROR (\%)} & \textbf{TIME (ms.)} & \textbf{COMP.}\\\hline

classic\_50  & 8.37 & 459 & 50\\\hline
classic\_100  & 4.13 & 940 & 100\\\hline
classic\_500  & 0.20 & 4700 & 500\\\hline
classic\_1000  & 0.20 & 9310 & 1000\\\hline
\hline
polyto\_3\_0.010  & 0.30 & \textbf{110} & \textbf{11}\\\hline
polyto\_3\_0.005  & \textbf{0.07} & 139 & 14\\\hline
polyto\_3\_0.003  & \textbf{0.07} & 140 & 14\\\hline
polyto\_4\_0.010  & 0.40 & 140 & 15\\\hline
polyto\_4\_0.005  & 0.20 & 149 & 16\\\hline
polyto\_4\_0.003  & 0.10 & 169 & 18\\\hline
polyto\_5\_0.010  & 0.30 & 150 & 16\\\hline
polyto\_5\_0.005  & \textbf{0.07} & 190 & 20\\\hline
polyto\_5\_0.003  & \textbf{0.07} & 179 & 20\\\hline
polyto\_6\_0.010  & 0.40 & 140 & 15\\\hline
polyto\_6\_0.005  & 0.10 & 179 & 19\\\hline
polyto\_6\_0.003  & 0.10 & 179 & 19\\\hline
polyto\_7\_0.010  & \textbf{0.0}7 & 190 & 21\\\hline
polyto\_7\_0.005  & \textbf{0.07} & 190 & 21\\\hline
polyto\_7\_0.003  & \textbf{0.07} & 200 & 21\\\hline

    \end{tabular}
    \end{table}

\begin{table}[!tb]
  \centering
  \caption{Fastest EER Computing Parameters For Each Modality}
  \label{tab:resspeedtot}
 \small
  \begin{tabular}{|l|l|c|c|c|c|}\hline
  \textbf{DB}&\textbf{LABEL} & \textbf{ERROR (\%)} & \textbf{TIME} & \textbf{COMP.}\\\hline
1& polyto\_3\_0.010  & 0.30 & 110 & 11\\\hline
2& polyto\_3\_0.010  & 0.05 & 50 & 5\\\hline
3& polyto\_6\_0.003  & 0.09 & 60 & 7\\\hline
4& polyto\_3\_0.010  & 0.14 & 89 & 10\\\hline
5& polyto\_4\_0.010  & 0.29 & 70 & 7\\\hline
    \end{tabular}
\end{table}

We can argue that the proposed method is better, both in terms of speed and precision error,
than the classical way of computing.
Based on the results of our dataset, the configuration using 3 steps and a
precision of 0.010 seems to be the best compromise between speed and precision.
We can now argue that the proposed EER computation will speed up genetic algorithms
using the EER as fitness function.

\subsubsection{Utility for Non-parametric Confidence Interval}

All the methods under all the implementations give similar confidence intervals.
We do not discuss on this point, because we are only interested in computation time.
Table~\ref{tab_computation_confidence} gives for each method, under each
implementation the mean value of the computation time for all the six different
biometric systems.
We can observe that, in average, the polytomous version seems far more faster than the other
methods (classic and whole).
The distributed implementation seems also more faster than the other
implementations (Parallel, Single).
Using the Kruskal-Wallis test at a confidence degree equals to $95\%$, the computation time of the
proposed method is significantly faster than both classical (\emph{p} value $=0.00651$) and whole (\emph{p} value $=0.004407$) schemes. There was no significant difference of computation time between classical and whole schemes (\emph{p} value $=0.8002$).

\begin{table}
\centering
\caption{Summary of the computation time in seconds. Time are averaged on all
the set of scores}
\label{tab_computation_confidence}
\begin{tabular}{|c|c|c|c||c|}\hline
& Polytomous & Classic & Whole & Mean \\ \hline
Distributed &  14.93 & 94.25 & 1009.28 & 372.82\\ \hline
Parallel    & 14.95 & 276.60 & 3154.32  & 1148.63 \\ \hline
Single      & 18.83 & 523.79 & 6733.68 & 2425.43\\ \hline          \hline
Mean        & 16.24 & 298.22 & 3632.43 & \\ \hline
\end{tabular}
\end{table}

%\todo[inline]{A toi de jouer mohamad. Par contre je comprend pas pourquoi le
%calcul distribué n'est pas plus rapide...}
%
%
%Interpretation -h statistic, pvalue)
%EER computing method dicho/classic/whole, kruskal
%(10.474523007856334, 0.0053147914608184374)
%dicho, classic (7.4034034034034022, 0.0065100595348351999)
%dicho, whole (8.1081081081081123, 0.0044067695476555343)
%whole, classic (0.064064064064069726, 0.80018413970643498)
%EER computing implementation distributd, parallel, single
%(1.0922558922558778, 0.57918812176233603)

%

\subsubsection{Discussion}
We have demonstrated the superiority of our EER estimation method against the classical method concerning the computation time.
However, the method stops when the required precision is obtained.
As the method is iterative, it is not parallelizable when we want only a simple EER.
However using a grid computing method greatly improves the computation of confidence intervals.

\section{Application to Multibiometrics Fusion Function Configuration}
We propose a biometric fusion system based on the generation of a fusion
function parametrized by a genetic algorithm and a fast method to compute the
EER (which is used as
fitness function)  in order to
speed the computing time of the genetic algorithm.

\subsection{Method}
We have tested three different kinds of score fusion methods which parameters are
automatically set by genetic algorithms~\cite{mitchell1998introduction}. These functions are presented in
(\ref{eq:ga1}),
(\ref{eq:ga2}) and
(\ref{eq:ga3}) where $n$ is the number of available
scores (\emph{i.e.}, the number of biometric systems involved in the fusion process),
$w_i$ the multiplication weight of score $i$, $s_i$ the score $i$ and
$x_i$ the weight of exponent of score $i$. (\ref{eq:ga1}) is the commonly used
weighted sum (note that in this version, the sum of the weights is not equal to
1), while the two others, to our knowledge, have never been
used in multibiometrics. We have empirically designed them in order to give more weights to higher scores.

\begin{equation}
  \label{eq:ga1}
  ga1 = \sum_{i=0}^{n}{w_i*s_i}
\end{equation}

\begin{equation}
  \label{eq:ga2}
  ga2 = \prod_{i=0}^{n}{s_i^{x_i}}
\end{equation}

\begin{equation}
  \label{eq:ga3}
  ga3 = \sum_{i=0}^{n}{w_i*s_i^{x_i}}
\end{equation}

The aim of the genetic algorithm is to optimize the parameters of each
function in order to obtain the best fusion function.
Each parameter (the $w_i$ and $x_i$) is stored in a chromosome of real
numbers. The fitness function is the same for the three genetic algorithms.
 It is processed in two steps:
\begin{itemize}
  \item
 \emph{fusion}: The generated function (\ref{eq:ga1}), (\ref{eq:ga2}) or
    (\ref{eq:ga3}) are evaluated on the whole set of scores;
  \item
 \emph{error computing}: The EER is computed on the result of the fusion. We
    use the polytomous version of computing in order to highly speed up the
    total computation time.
\end{itemize}

\subsection{Experimental Protocol}
\subsubsection{Design of Fusion Functions}
Table~\ref{tab:gaparams} presents the parameters of the genetic algorithms.
The genetic algorithms have been trained on a learning set composed of half of the
intrascores and half of the interscores  of a database and they have been verified with a
validation set composed of the others scores. The three databases have been used
separately.

\begin{table}[!tb]
\caption{Configuration of the Genetic Algorithms}
\centering
\label{tab:gaparams}
\small
\begin{tabular}{|l|p{6cm}|}\hline
\textbf{Parameter} & \textbf{Value}\\\hline\hline
Population & 5000  \\\hline
Generations & 500   \\\hline
Chromosome signification & weights and powers of the fusion functions \\\hline
Chromosome values interval & $[-10;10]$ \\\hline
Fitness    & polytomous EER on the generated function\\\hline
Selection  &  normalized gemetric selection (probability of 0.9)\\\hline
Mutation &

boundary,
multi non uniform,
non uniform,
uniform

 \\\hline
Cross-over & Heuristic Crossover\\\hline
%Copy rate & \color{red}??\color{black} \\\hline
Elitism & True \\\hline
\end{tabular}
\end{table}

The generated functions are compared to three methods of the state of the art:
$sum$, $mul$ and $min$, they have been explored
in~\cite{jain2005snm,Kittler1998CC}.
 Table~\ref{tab:initperf} presents, for each
database, the EER value of each of its biometric method (noted $sn$ for method $n$),
as well as the performance of the fusion functions of the state of the art.
\begin{table}[!tb]
  \caption{Performance (EER) of the Biometric Systems ($s1$, $s2$, $s3$, $s4$),
and the State Of The Art Fusion Functions ($sum$,$min$,$mul$) on the Three Databases}
  \label{tab:initperf}\centering
\small
  \begin{tabular}{|ll|c|c|}\hline
    \multicolumn{2}{|l|}{\textbf{Method}} & \textbf{Learning} & \textbf{Validation}\\\hline
    \multicolumn{4}{|c|}{BANCA}\\\hline

\multirow{4}{*}{Biometric systems}
&$s1$&0.0310&0.0438\\%\hline
&$s2$&0.0680&0.1154\\%\hline
&$s3$&0.0824&0.0897\\%\hline
&$s4$&0.0974&0.0732\\\hline
\multirow{3}{*}{State of the art fusion}
&$sum$&\textbf{0.0128}&\textbf{0.0128}\\%\hline
&$min$&0.0385&0.0438\\%\hline
&$mul$&\textbf{0.0128}&\textbf{0.0128}\\\hline

    \multicolumn{4}{|c|}{BSSR1}\\\hline
\multirow{4}{*}{Biometric systems}
&$s1$&0.0425&0.0430\\%\hline
&$s2$&0.0553&0.0620\\%\hline
&$s3$&0.0861&0.0841\\%\hline
&$s4$&0.0511&0.0454\\\hline
\multirow{3}{*}{State of the art fusion}
&$sum$&\textbf{0.0116}&\textbf{0.0070}\\%\hline
&$min$&0.0436&0.0504\\%\hline
&$mul$&0.0117&\textbf{0.0070}\\\hline

    \multicolumn{4}{|c|}{PRIVATE}\\\hline
\multirow{4}{*}{Biometric systems}
&$s1$&0.1161&0.1153\\%\hline
&$s2$&0.1522&0.1569\\%\hline
&$s3$&0.0603& 0.0621\\%\hline
&$s4$&0.2815&0.3143\\\hline
\multirow{3}{*}{State of the art fusion}
&$sum$&0.0256&\textbf{0.0278}\\%\hline
&$min$&0.1397&0.1471\\%\hline
&$mul$&\textbf{0.0252}&0.0281\\\hline
\end{tabular}
\end{table}
We can see that biometric methods from PRIVATE have more biometric verification
errors than the
ones of the other databases. Using the Kruskal-Wallis test, the $sum$ (\emph{p} value $=0.0038$)
and $mul$ (\emph{p} value $=0.0038$) operators outperformed the $min$ operator. There was no
significant difference (\emph{p} value $=0.935$) between both operators $sum$ and $mul$ operators.\\

\subsubsection{Magnitude Of the Gain in Computation Time}
We also want to prove that using the proposed EER computation method improves the
computation time of the genetic algorithm run.
To do that, the previously described process has been repeated two times:
\begin{itemize}
 \item
Using the proposed EER computing method with the following
configuration: 5 steps and stop at a precision of 0.01.
 \item
 Using the classical EER computing  method with 100 steps.
\end{itemize}~
The total computation time is saved in order to compare the speed of the two
systems. These tests have been done on a Pentium IV machine with 512 Mo of RAM
with the Matlab programming language.\\

\subsection{Experimental Results}
\subsubsection{Design of Fusion Functions}
\begin{figure*}[!tb]
{\includegraphics[width=0.9\linewidth]{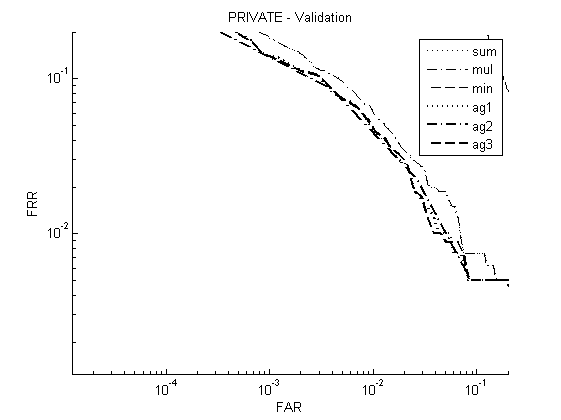}}
\caption{ROC Curve Of The Generated Multibiometrics Fusion Functions on the
Validation Set of the Private dataset}
\label{fig:resfusion}
\end{figure*}

%Table~\ref{tab:gares} presents the generated multibiometric fusion functions
%with the learning set.
%Their global performances are presented in Figure~\ref{fig:resfusion}, by displaying
%their ROC curve computed on the validation set (the results of the $sum$, $min$
%and $mul$ functions are also represented) while the EER of each set is presented
%in Table~\ref{tab:globalres}.
The EER of each generated function of each database is presented in
Table~\ref{tab:globalres} for the learning and validation sets, while
Figure~\ref{fig:resfusion} presents there ROC curve on the validation set.
We can see that the proposed generated functions are all globally better than the ones from the
state of the art (by comparison with Table~\ref{tab:initperf}) and the obtained EER is always
better than the ones of the $sum$
and $mul$. The two new fusion functions ((\ref{eq:ga2}) and (\ref{eq:ga3})) give
similar or better results than the weighted sum (\ref{eq:ga1}).
We do not observe over-fitting problems: the results are promising both on the
learning and validation sets.
%
%
%\begin{comment}
\begin{table*}[!tb]
\centering
\caption{Configurations Of The Three Weighted Functions For Each Database}
\label{tab:gares}
\small
\begin{tabular}{|l|l|}\hline
\multicolumn{2}{|c|}{BANCA}\\\hline
GA & configuration \\\hline
ga1 &$8.7229*s_0 + 2.3092*s_1 2.0626*s_2 + 2.9687*s_3   $ \\\hline
ga2 &$s_0^{7.4721}*s_1^{1.8091}*s_2^{2.1255}*s_3^{0.8874}   $ \\\hline
ga3 &$-2.3079*s_0^{-6.0105} + -8.5217*s_1^{-3.1367} +  -8.6644*s_2^{-3.5730} +
-7.0890*s_3^{7.4634}   $ \\\hline

\hline
\multicolumn{2}{|c|}{BSSR1}\\\hline
GA & configuration \\\hline
ga1 & $2.3270*s_0 +    0.8790*s_1  +  0.3661*s_2 +   9.4978*s_3   $\\\hline
ga2 & $s_0^{2.0650} *  s_1^{0.5660}  *  s_2^{1.6168}  * s_3^{9.1864}   $\\\hline
ga3 & $5.7285*s_0^{4.6227} +   4.2471*s_1^{6.8192} +    9.7541*s_2^{6.7588} +  5.9431*s_3^{0.9251}   $\\\hline

\hline
\multicolumn{2}{|c|}{PRIVATE}\\\hline
GA & configuration \\\hline
ga1 & $6.7755*s_0 +    2.3841*s_1  +  5.9128*s_2 +    2.6919*s_3   $\\\hline
ga2 & $ s_0^{6.2215} *   s_1^{4.1538} * s_2^{6.6853} * s_3^{3.9254}   $\\\hline
ga3 & $4.8647*s_0^{1.6977} +   8.3564*s_1^{5.9125} + 4.7450*s_2^{2.2407} +   2.0707*s_3^{0.7681}   $\\\hline

\end{tabular}
\end{table*}
%\end{comment}
%
%
\begin{figure*}[!tb]
\centering
{\includegraphics[width=0.9\linewidth]{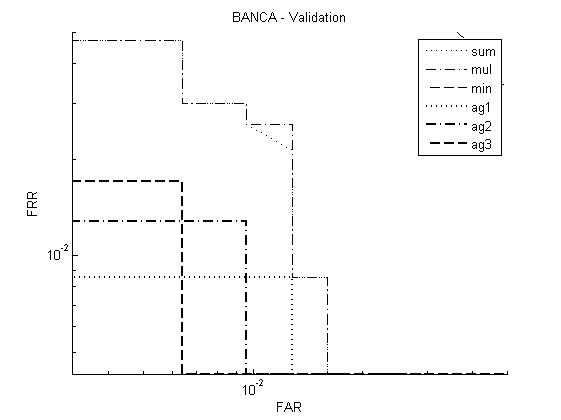}}
\caption{ROC Curve Of The Generated Multibiometrics Fusion Functions on the
Validation Set of the BANCA dataset}

\end{figure*}
\begin{figure*}[!tb]
{\includegraphics[width=0.9\linewidth]{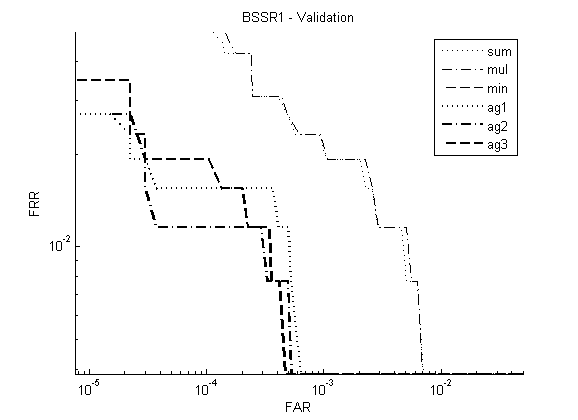}}
\caption{ROC Curve Of The Generated Multibiometrics Fusion Functions on the
Validation Set of the BSSR1 dataset}

\end{figure*}
\begin{table}[!tb]
\centering
\caption{
EER For Training and Validation
Sets And Computation Time Gain By Using Our EER Computation Method}
\label{tab:globalres}
\small
\begin{tabular}{|l|c|c||c|}\hline
\textbf{Function}& \textbf{Train EER} & \textbf{Test EER} &
\textbf{Gain (\%)}\\\hline
\multicolumn{4}{|c|}{BANCA}\\\hline
(\ref{eq:ga1}): ga1 & \textbf{0.0032}   & 0.0091    &\textbf{61.29} \\\hline
(\ref{eq:ga2}): ga2 & \textbf{0.0032}   & 0.0091    &41.84 \\\hline
(\ref{eq:ga3}): ga3 & 0.0037    & \textbf{0.0053}   &43 \\\hline

\multicolumn{4}{|c|}{BSSR1}\\\hline
(\ref{eq:ga1}): ga1  &  0.000596    & \textbf{0.0038}   & \textbf{78.32} \\\hline
(\ref{eq:ga2}): ga2 & \textbf{0.000532} & \textbf{0.0038}   & 64.77 \\\hline
(\ref{eq:ga3}): ga3 & 0.000626  & \textbf{0.0038}   & 28.49  \\\hline

\multicolumn{4}{|c|}{PRIVATE}\\\hline
(\ref{eq:ga1}): ga1  &  0.019899&   0.0241  & \textbf{77.66} \\\hline
(\ref{eq:ga2}): ga2 & \textbf{0.019653} & 0.0244    & 46.5 \\\hline
(\ref{eq:ga3}): ga3 & 0.020152  & \textbf{0.0217}   & 55.03 \\\hline

\end{tabular}
\end{table}
We also could expect to obtain even better performance by using more individuals
or more generations in the genetic algorithm process, but, in this case, timing
computation would become too much important. Their will always be a tradeoff between security
(biometric performance) and computation speed (genetic algorithm performance).
By the way, the best individuals were provided in the first 10 generations, and
several runs give approximately the same results, so we
may already be in a global minima.

As a conclusion of this part, we increased the performance of multibiometrics
systems given the state of the art by reducing errors of 58\% for BANCA, 45\%
for BSSR1 and 22\% for PRIVATE.
\subsubsection{Magnitude Of the Gain in Computation Time}
Table~\ref{tab:globalres} presents a summary of the performance of the generated
methods both in term of EER and timing computing improvement.
The column gain presents the improvement of timing computation between the proposed EER
polytomous computation time and the classical one in 100 steps.

We can observe that, in all the cases, the proposed computation methods outperform the
classical one (which is not its slowest version). We can see that this improvement
depends both on the cardinal of the set of scores and the function to evaluate:
there are better improvements for (\ref{eq:ga1}). The best gain is about 78\%
while the smallest is about 28\%.

\subsubsection{Discussion}
Once again, we can observe the interest of our EER estimation method which allows to obtain results far quickly.
We can note that the generated function all generate a monotonically decreasing ROC curve which allows to use our method.
If the ROC curve does not present this shape, we would be unable to obtain the
estimated EER (such drawbacks as been experimented using genetic
programming~\cite{giot2011genetic} instead of genetic algorithms).

\section{Conclusion and Perspectives}
The contribution of this paper is twofold:
a fast approximated EER computing method (associated to its confidence interval), and
two score fusion functions having to be parametrized thanks to genetic
algorithms.
Using these two contributions together allows
to speed up the computation time of the genetic algorithm because its fitness function consists
on computing the EER (thus, allows to use a bigger population).

The fast EER computing method has been validated on five different biometric
systems and compared to the classical way.
Experimental results showed the benefit
of the proposed method, in terms of precision of the EER value and timing
computation.

The score fusion functions have been validated on three significant
multibiometrics databases (two reals and one chimerical) having a different number of scores.
The fusion functions
parametrized by genetic algorithm always outperform simple state of the art
simple functions ($sum$, $min$, $mul$), and, the two new fusion functions have given
better are equal results than the simple weighted sum.
Using the proposed fast EER computing
method also considerably speed up the timing computation of the genetic
algorithms.
These better results imply that the multibiometrics system has a better security
(fewer impostors can be accepted) and is more pleasant to be used (fewer genuine
users can be rejected).

One limitation of the proposed method is related to the shape of the ROC curve and the
atended precision wanted.
In some cases, the method is unable to get the EER at the wanted precision, and,
is not able to return the result (we did not encounter this case in these
experiments).

Our next research will focus on the use of different evolutionary algorithms in
order to generate other kind of complex functions allowing to get better
results.

\section*{Acknowledgements}
Work has been with financial support of the ANR ASAP project, the Lower Normandy
French Region and the French research ministry.

%\bibliography{dicho_eer}

\end{document}